# Manipulating Google Scholar Citations and Google Scholar Metrics: simple, easy and tempting

Emilio Delgado López-Cózar[1], Nicolás Robinson-García[1] y Daniel Torres-Salinas[2]
EC3: Evaluación de la Ciencia y de la Comunicación Científica
[1]Universidad de Granada
[2]Universidad de Navarra
edelgado@ugr.es; elrobin@ugr.es; torressalinas@gmail.com

**ABSTRACT**
**The launch of Google Scholar Citations and Google Scholar Metrics may cause a revolution in the research evaluation field as it places within every researcher's reach tools that allow them to measure their output. However, the data and bibliometric indicators offered by Google's products can be easily manipulated. In order to alert the research community, we present an experiment in which we manipulate the Google Citations' profiles of a research group through the creation of false documents that cite their documents, and consequently, the journals in which they have published, modifying their H-index. For this purpose we created six documents authored by a faked author and we uploaded them to a researcher's personal website under the University of Granada's domain. The result of the experiment meant an increase of 774 citations in 129 papers (six citations per paper) increasing the authors and journals' H-index . We analyse the malicious effect this type of practices can cause to Google Scholar Citations and Google Scholar Metrics. Finally, we conclude with several deliberations over the effects these malpractices may have and the lack of control these tools offer.**

## 1. INTRODUCTION

If the launch of Google Scholar in 2004 (a novel search engine focused on retrieving any type of academic material along with its citations) meant a revolution in the scientific information market allowing universal and free access to all documents available in the web, the launch of Google Scholar Citations (hereafter GS Citations) (Cabezas-Clavijo & Torres-Salinas, 2012) and Google Scholar Metrics (hereafter GS Metrics) (Cabezas-Clavijo & Delgado López-Cózar, 2012) may well be a historical milestone for the globalization and democratisation of research evaluation (Butler 2011). As well as constituting a new threat to the traditional bibliographic databases and bibliometric indexes offered by Thomson Reuters (Web of Science and JCR) and Elsevier (Scopus and SJR), ending with their monopoly and becoming a serious competitor; Google Scholar's new products project a future landscape with ethical and sociological dilemmas that may entail serious consequences in the world of science and research evaluation.

Without considering the technical and methodological problems that the Google Scholar products have which are currently under study (Jacsó, 2008, 2011; Wouters y Costas, 2012; Aguillo, 2012; Cabezas-Clavijo y Delgado López-Cózar, 2012; Torres-Salinas, Ruiz-Pérez y Delgado López-Cózar, 2009) and which will be presumably solved in a near future, its irruption ends with all kinds of scientific control, becoming a new challenge to the bibliometric community. Since the moment Google Scholar automatically retrieves, indexes and stores any type of scientific material uploaded by an author without any previous external control (repositories are only a technical filter as they do not assure a revision of the content), it allows unprincipled people to manipulate their output, impacting directly on their bibliometric performance.

Because this type of behaviour by which one modifies its output and impact through intentional and unrestrained self-citation is not uncommon, we consider necessary to analyse thoroughly Google's capacity to detect the manipulation of data.

This study continues the research line started by Labbé (2010). In his experiment he transformed a faked researcher called Ike Antkare ( *'I can't care'*) into the most prolific researcher in history. However, in this case we will enquire over the most dangerous aspects of gaming tools aimed at evaluating researchers and the malicious effects they can have on researchers' behaviour. Therefore our aim is to demonstrate how easily anyone can manipulate Google Scholar's tools. But, contrarily to Labbé, we will not emphasize the technical aspects of such gaming, but its sociological dimension, focusing on the enormous temptation these tools can have for researchers and journals' editors, eager to increase their impact. In order to do so, we will show how the bibliometric profiles of researchers and journals can be modified simultaneously in the easiest way possible: by uploading faked documents on our personal website citing the whole production of a research group. It is not necessary to use any type of software for creating faked documents: you only need to copy and paste the same text over and over again and upload the resulting documents in a webpage under an institutional domain. We will also analyse Google's capacity to detect retracted documents and delete their bibliographic records along with the citations they make.

This type of studies by which false documents are created in order to evidence defects, biases or errors committed by authors have been conducted many times, especially in the research evaluation field. The reader is referred to the works of Peters & Ceci (1990), Epstein (1990), Sokal (1996, 1997) or Baxt et al. (1998) when demonstrating the deficiencies of the peer review method as an objective, reliable and valid control tool when filtering scientific papers. Or Scigen[1], a programme created by three students from the MIT for generating random papers in the Computer Science field including graphs, figures and references. All of these works raised an intense debate within the research community.

Therefore, this paper is structured as follows. Firstly we described the methodology followed; how were the false documents created and where were they uploaded. Then we show the effect they had on the bibliometric profiles of the researchers who received the citations and we emulate the effect these citations would have had on the journals affected if GS Metrics was updated regularly. We analyse the technical effects and the dangers these tools entail for evaluating research. Finally we conclude emphasizing their strengths and we end with some concluding remarks.

---

[1] http://pdos.csail.mit.edu/scigen/

## 2. MANIPULATING DATA: THE GOOGLE SCHOLAR EXPERIMENT

In order to analyse GS Citations' capacity to discriminate academic works from those which aren't and test the grade of difficulty for manipulating output and citations in Google Scholar and its bibliometric tools (GS Citations and Metrics), we created false documents referencing the whole research production of the EC3 research group (Science and Scientific Communication Evaluation) available at http://ec3.ugr.es in the easiest possible way. This way we intend to show how anyone can manipulate its output and citations in GS Citations.

**Figure 1. Fake documents authored by the non-existent researcher MA Pantani-Contador**

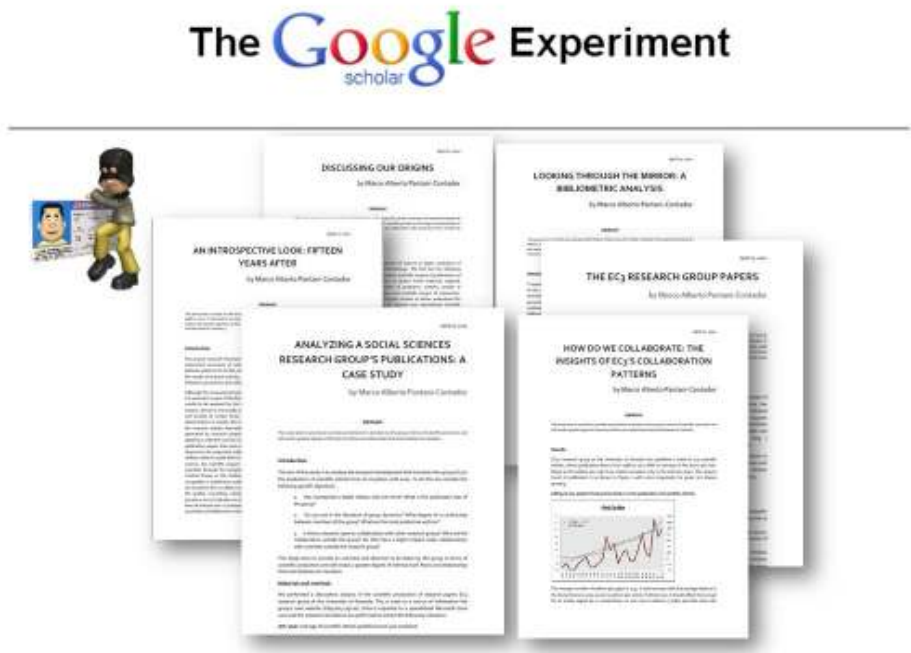

Following the example set by Labbé (2010), we created a false researcher named Marco Alberto Pantani-Contador, making reference to the great fraud the Italian cyclist became at the end and the accidental causes that deprived the Spanish cyclist from winning the Tour. Thus, Pantani-Contador authored six documents (figure 1) which did not intend to be considered as research papers but working papers. In a process that lasted less than a half day's work, we draft a small text, copied and pasted some more from the EC3 research group's website, included several graphs and figures, translated it automatically into English using Google Translate and divided it into six documents. Each document referenced 129 papers authored by at least one member of the EC3 research group according to their website http://ec3.ugr.es. That is, we expected a total increase of 774 citations.

Afterwards, we created a webpage in html under the University of Granada domain including references to the false papers and linking to their full-text, expecting Google Scholar to index their content. We excluded other services such as institutional or subject-based repositories as they are not obliged to undertake any bibliographic control rather than a formal one (Delgado López-Cózar, 2012) and we did not aim at bypassing their filters.

The false documents were uploaded on 17 April, 2012. Presumably because it was a personal website and not a repository, Google indexed these documents nearly a month after they were uploaded, on 12 May, 2012. At that time the members of the research group used as case study, including the authors of this paper, received an alert from GS Citations pointing out that someone called MA Pantani-Contador had cited their output. The citation explosion was thrilling, especially in the case of the youngest researchers where their citation rates were multiplied by six, notoriously increasing in size their profiles.

**Figure 2. Citations increase for the authors of this paper**

**Emilio Delgado López-Cózar**

| | WHOLE PERIOD | | | SINCE 2007 | | |
|---|---|---|---|---|---|---|
| | BEFORE the experiment | AFTER the experiment | | BEFORE the experiment | AFTER the experiment | |
| Citations | 862 | 1297 | + 435 | 560 | 995 | + 435 |
| H-Index | 15 | 17 | + 2 | 10 | 15 | + 5 |
| i10-Index | 20 | 40 | + 20 | 11 | 33 | + 22 |

**Nicolás Robinson-García**

| | WHOLE PERIOD | | | SINCE 2007 | | |
|---|---|---|---|---|---|---|
| | BEFORE the experiment | AFTER the experiment | | BEFORE the experiment | AFTER the experiment | |
| Citations | 4 | 29 | + 25 | 4 | 29 | + 25 |
| H-Index | 1 | 4 | + 3 | 1 | 4 | + 3 |
| i10-Index | 0 | 0 | 0 | 0 | 0 | 0 |

**Daniel Torres-Salinas**

| | WHOLE PERIOD | | | SINCE 2007 | | |
|---|---|---|---|---|---|---|
| | BEFORE the experiment | AFTER the experiment | | BEFORE the experiment | AFTER the experiment | |
| Citations | 227 | 409 | + 182 | 226 | 408 | + 182 |
| H-Index | 9 | 11 | + 2 | 9 | 11 | + 2 |
| i10-Index | 7 | 17 | + 10 | 7 | 17 | + 10 |

Figure 2 shows the increase of citations the authors experienced. Obviously, the number of citations by author varies depending on the number of publications each member of the research group had as well as the inclusion of real citations received during the study period. Thus, the greatest increase is for the less-cited author, Robinson-Garcia, who multiplies by 7.25 the number of citations received, while Torres-Salinas doubles it and Delgado López-Cózar experiences an increase of 1.5. We also note the effect on the H-index of each researcher. While the most significant increase is perceived in the less prolific profile, the variation for the other two others is much more moderate, illustrating the stability of the indicator. Note how in Torres-Salinas' case, where the number of citations is doubled, how the H-index only increases by two. On the other hand, we observe how the i10-index is much more sensitive to changes. In Torres-Salinas' case, the increase goes from 7 to 17, and in Delgado López-Cózar's case it triples for the last five years, going from 11 to 33.

**Figure 3. Effects on the manipulation of the citations in one of the authors**

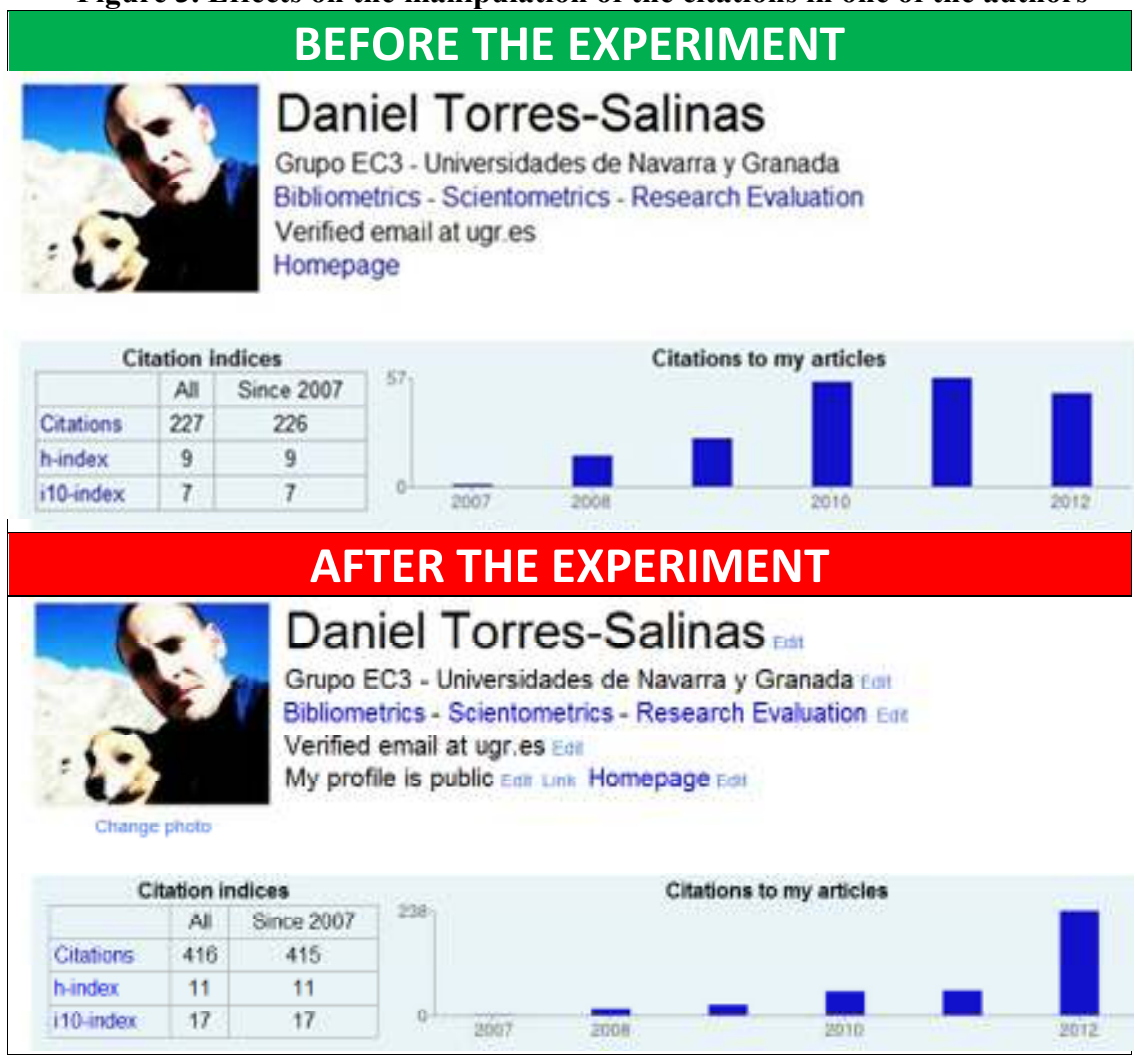

Also, it is interesting to analyse the effect this citation increase may have on the h-index for journals indexed in GS Metrics. For this, we have considered the two journals in which the members of the research group have published more papers and therefore, more sensitive to be manipulated. These are El Profesional de la Información with 30 papers published in this journal and Revista Española de Documentación Científica, with 33 papers. In table 1 we show the H-indexes for El Profesional de la Información and Revista Española de Documentación Científica according to Google and the increase it would have had if the citations emitted by Pantani-Contador had been included. We must alert the reader that this tool, contrarily to the rest of Google's products, is not automatically updated and that data displayed dates to the day of its launch, that is, 1 April, 2012 (Cabezas-Clavijo & Delgado López-Cózar, 2012). We observe that El Profesional de la Información would be the one which would have been more influenced, as seven papers would surpass the 12 citations threshold increasing its H-index and ascending in the ranking for journals in Spanish language from position 20 to position 5. Revista Española de Documentación Científica would slightly modify its position, as only one article surpasses the 9 citations threshold that influence its h-index. Even so and due to the high number of journals with its same h-index, it would go up from position 74 to 54.

Table 1. Effect of the manipulation of citations over journals

| Journal | H-Index (GS Metrics) | Art > H-Index citation threshold | Manipulated H-Index |
|---|---|---|---|
| El Profesional de la Información | 12 | 7 | 19 |
| Revista Española de Documentación Científica | 9 | 1 | 10 |

After proving the vulnerability of Google's products when including false documents and showing its effect at the researcher-level and journal-level, on 17 May, 2012 we deleted the false documents and webpage in order to see if Google Scholar would delete the records and the citations received according to GS Citations. However, until this date (29 May) and 17 days after they were removed from the Internet, no modifications have been made whatsoever. The records of the authored documents by our faked researcher are still available when searching its production and, despite being broken links, there is a version of the documents saved by Google.

## 3. TECHNICAL CONSIDERATIONS

The results of our experiment show how easy and simple it is to modify the citation profiles offered by GS Citations. This exposes the vulnerability of the product if editors and researchers are tempted to do "citations engineering" and modify their H-index by excessively self-citing their papers or, in a most refined way, sending citations only to the hot zone of their publications, that is, those which can influence this indicator. In the case of El Profesional de la Información, there are 16 documents with 10 to 12 citations for the time period analysed by GS Metrics (2007-2011). These could modify the journal's position by receiving 1 to 3 citations more.

Back to more technical issues, firstly, we must emphasize how easy it is to manipulate, not just output, - as previously shown by Labbé (2010), - but also citations. This raises serious concerns over the limitations of Google Scholar when discriminating faked documents. Although Google Scholar is only meant to index and retrieve all kinds of academic material in its widest sense, the inclusion of GS Citations and GS Metrics, which are evaluating tools, must include some filters and monitoring tools as well as the establishment of a set of more rigid criteria for indexing documents. Google Scholar offers access to a wide range of document types, becoming a much more attractive database. Not just because of its "magic formula" for retrieving and ranking results, but because of the richness of the data it handles. However, leaving such a controlled environment as journals leads to many dangers in the research evaluation world.

On the other hand, it is interesting to observe the stability of the h-Index when affecting experienced researchers, even if the number of citations is doubled. This may bring a sense of relief, however, unfortunately there are many ways for manipulating this indicator through self-citation (Bartneck and Kokkelmans, 2011). Also, regarding journals and the most likely future update of GS Metrics, which was included on Google Scholar`s homepage a few days ago, devious editors can easily modify their journals' H-index. Also, we observe how notable is the variation of the i10-Index, especially for experienced researchers.

Regarding the effect these malpractices may have over the rankings presented by Google, it would obviously be significant, especially for those journals with small figures, on which the slightest variation can have a great impact on their performance.

The impossibility of editing citations in GS Citations pointing out the wrong ones and indicating those which have not been detected, highlights this shortcoming, therefore we alert as others have done (Cabezas-Clavijo & Torres-Salinas, 2012) of the limitations of these tools when used for bibliometric purposes. The last part of the experiment will be to see if the records of the deleted documents will be eventually removed from Google Scholar along with the references. This has not still happened and, if it doesn't occur, it will emphasize an important shortcoming the general search engine also has, its impossibility to exercise our "right to be forgotten" (Gómez, 2011).

**Figure 4. Results from Google Scholar**

Now, it is important to emphasize the visibility these tools offer and the transparency they seem to show, facilitating the detection of these practices by the community, as we have witnessed over the elaboration of this experiment. Many of the co-authors affected by the malpractices of devious Pantani-Contador detected his reproachable behaviour and enquired over the issue.

On the other side, it is interesting to see how papers published over the same template are indexed differently by Google. This shows once again, the lack of normalization it has. Therefore we see naming variations over the six false documents uploaded (figure 4).

## 3. FINAL THOUGHTS AND CONCLUSIONS

Even if we have previously argued in favour of Google Scholar as a research evaluation tool minimizing its biases and technical and methodological issues (Cabezas-Clavijo, Delgado López-Cózar, 2012), in this paper we alert the research community over how easy it is to manipulate data and bibliometric indicators. Switching from a controlled environment where the production, dissemination and evaluation of scientific knowledge is monitored (even accepting all the shortcomings of peer review) to a environment that lacks of any kind of control rather than researchers' consciousness is a radical novelty that encounters many dangers. (Table 2).

**Table 2. Control measures in the traditional model vs. Google Scholar's products**

| Traditional model | Google Scholar's tools |
|---|---|
| Databases select journals to be indexed | It indexes any document belonging to an academic domain |
| Journals select papers to be published | Any indexed document type emits and receives citations |
| There is a control between citing and cited documents<br><br>Fraudulent behaviours are persecuted | It is not possible to alert over fraudulent behaviours or citation errors |

Putting on researchers' hand, which are humans, the tools that allow manipulating output and citations may have unforeseen consequences or make these tools useless. The lack of control that characterises these tools is their main strength but also their weakness. It is so easy to manipulate GS Citations that anyone can emulate Ike Antkare and become the most productive and influential researcher in its specialty. Let alone editors, if GS Metrics is finally incorporated, they can be tempted to use unethical techniques to increase the impact of their journals.

These free and accessible products, do not only awaken the Narcissus within researchers (Wouters; Costas, 2012), but can unleash malpractices aiming at manipulating the orientation and meaning of numbers as a consequence of the ever growing pressure for publishing fuelled by the research evaluation exercises of each country. There are many cases of editors' frauds where they manipulate through editorial policies researchers' behaviours in order to increase the impact factor (see e.g., Falagas & Alexiou, 2008).

Many journals are excluded every year from the Web of Science because of their fraudulent behaviour[2]. There are many examples, such as the one reported by Dimitrov et al. (2010) with the resounding case of revista *Acta Crystallographica A* which surprised everyone when increasing its impact factor from 2,38 to 49,93 in a year. It seemed that from the 5966 citations received in 2009 by the 72 papers published in 2008, 5624 belonged just to one article. This paper was in fact responsible of such an anomalous behaviour. Another example can be found in Opatrný (2008).

Currently there are no controlling or filtering systems for avoiding fraud rather than researchers' ethical values. In this sense, we must point out the role of institutions such as the Committee on Publication Ethics[3] and other similar organizations devoted to pursuing fraud within the traditional research communication model, that is, journals. We may be witnessing a new revolution of the scientific communication model and it may be just a matter of time to see other similar organization working in this new environment. For our part, we conclude our experiment and we await patiently the retraction of our inexistent researcher by Google, following our example and deleting the faked citations from our profiles. Google's effort on the creation of new evaluation tools forecasts many changes in the research evaluation world. Not just because these tools are cost-free, but because of their great coverage, immediacy and ease of use. We will just have to wait to see which path will Google follow in their attempt to put a stop to those numbers that are devouring science (Monastersky 2005).

## SUPLEMENTARY MATERIAL

More information is available http://www.ugr.es/~elrobin/pantani.html.

---

[2] http://admin-apps.webofknowledge.com/JCR/static_html/notices/notices.htm
[3] http://publicationethics.org/